\documentclass[9pt,twocolumn,twoside]{osajnl}

\journal{josaa} 

\setboolean{shortarticle}{false} 

\title{Measurement of thickness of thin film by fitting to the intensity profile of Fresnel diffraction from a nano phase step}

\author[1,2,3,4,*]{Ali Motazedifard}
\author[1,5]{S. Dehbod}
\author[1,2]{A. Salehpour}

\affil[1]{Kahroba Technology Company, University of Tehran Science and Technology Park, Tehran 14399-55961, Iran }
\affil[2]{Department of Physics, College of Science, University of Tehran, Kargar Shomali Ave, Tehran 14399-55961, Iran}
\affil[3]{Quantum Optics Group, Department of Physics, Faculty of Science, University of Isfahan, Hezar Jerib, 81746-73441, Isfahan, Iran}
\affil[4]{School of physics, Institute for Research in Fundamental Sciences (IPM), Tehran 19395-5531, Iran}
\affil[5]{Department of Mathematics and Computer Science, Amirkabir University of Technology, Tehran, Iran}

\affil[*]{Corresponding author: motazedifard.ali@gmail.com}

\dates{Compiled \today}

\ociscodes{120.3940, 260.0260, 050.1940}

\doi{}

\begin{abstract}
Diffraction of light beams from the phase steps due to the abrupt/sharp changes in the boundary of step leads to Fresnel fringes whose visibility and intensity profile depend on the change of the step height or light incident angle. The visibility has been utilized in measurements of different physical quantities. In this paper, for the first time to our knowledge by introducing the fitting method as a fast method we show that by fitting the theoretical intensity distributions on the experimental intensity profiles of the light diffracted from a step at different incident angles, one can specify the step height with few nanometers precision. In addition, we show that this approach provides accurate film thickness in a broad range of thicknesses using modest instrumentation. Furthermore, based on Fresnel diffraction from phase step we have manufactured and trademarked an optical device for measuring thickness of thin films.

\end{abstract}

\setboolean{displaycopyright}{true}

\begin{document}

\maketitle
\thispagestyle{fancy}
\ifthenelse{\boolean{shortarticle}}{\abscontent}{}

\section{Introduction}
The Film thickness is an important parameter in many fields such as thin film physics, micro-electronics, optics, and industry. Therefore, a large number of techniques have been introduced for film thickness measurement \cite{1,pedroti,2,3,4,5,6,7,8}. Among them, optical techniques, particularly those based on interferometry and ellipsometry have been widely used because of their reliable results \cite{9,10,11,12,13,14,15}. Ellipsometry is more appealing because it can measure both the film thickness and the optical constants of the sample \cite{16,17}.

In this paper, for the first time to our knowledge, we introduce a new technique based on the intensity profile fitting of Fresnel diffraction (FD) of the light beam from the phase step (PS) which help us to measure film thickness with high precision in a wide range. We show both theoretically and experimentally that our new method which is based on fitting the theoretical intensity distributions on the experimental intensity profiles of the light beam diffracted from a PS at different incident angles has many advantages in comparison with the old method of using the linear part of the visibility curve \cite{18,19,20,21,22}. Among these advantages we can emphasize the high accuracy of the step height measurement with a few nanometers precision which is in good agreement with the results obtained from the visibility technique \cite{22} or others such as interferometry, AFM (atomic force macroscopic) and Profilometer. Our technique is very reliable and precise compared with the techniques based on conventional interferometry or ellipsometry because of their high sensitivity to the displacements while FD from PS is not sensitive to the displacement or vibration. Also, it can be applied easily with modest optical instrumentation. Furthermore, we have manufactured and trademarked a device for thickness measurement.

The paper is organized as follows. In Sec.~\ref{theory}, we describe the theoretical formulation of 1D FD from PS in reflection mode. Also, in this section we theoretically introduce the fitting method based on FD from PS and explain its mechanism by considering a simulation example. In Sec.~\ref{experiment}, we describe the experimental procedure and our results. Furthermore, we compare our results to AFM, interferometry, and profilometry and a VLSI standard step height. Finally, in Sec.~\ref{con} we summarize our conclusions, discuss the potentials of our technique and point out some outlooks.

\section{Theoretical approach} \label{theory}
\begin{figure}
	\centering
	\includegraphics[width=8cm]{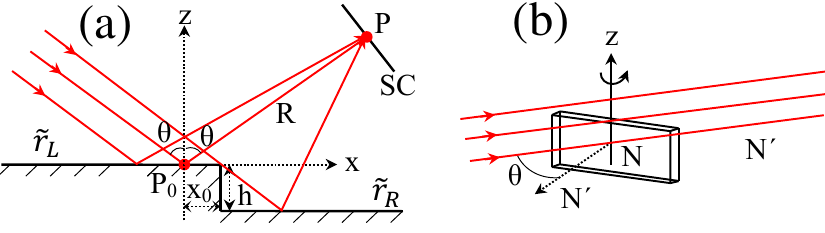}
	\caption{(Color online) (a) The geometry used to calculate the intensity profile on the diffraction pattern of light diffracted from a 1D phase step in reflection mode. (b) A phase step in transmission mode; as light passes through the boundary region of a transparent plate immersed in a transparent medium experiences a sharp change in phase. $ \tilde r_L $ and $ \tilde r_R $ are respectively the complex reflection coefficients of the left and the right sides of the step.}
	\label{fig1}
\end{figure}

\subsection{Theoretical formulation of FD from a phase step}

As a coherent beam of light strikes a physical step of height $ h $, see Fig.~\ref{fig1}(a), the reflected wave experiences a discontinuity in its phase that leads to intensity redistribution on a screen perpendicular to the reflected beam. Also, a similar effect is observed as a coherent light beam passes through the boundary region of a transparent plate immersed in a transparent medium, see Fig.~\ref{fig1}(b). In this case the discontinuity in the refractive index at the plate boundary causes a sharp/abrupt change in the phase. The intensity redistribution in both cases can be calculated by applying the Fresnel-Kirchhoff integrals \cite{18,19}. The corresponding calculations lead to the following expression for the intensity at point $ P $ in Fig.~\ref{fig1}(a) \cite{18,19};
\begin{eqnarray} \label{Ip}
&& \!\!\!\!\!\!\!\!\! {{I}_{(P)}}={{I}_{0}}{{r}_{L}}{{r}_{R}}\left[ \left(0.5-(C_{0}^{2}+S_{0}^{2})\right)\cos \varphi -({{C}_{0}}-{{S}_{0}})\sin \varphi  \right] \nonumber  \\
&& \quad +\frac{{{I}_{0}}}{2}\left[ \left(0.5+(C_{0}^{2}+S_{0}^{2})\right)(r_{L}^{2}+r_{R}^{2})+({{C}_{0}}+{{S}_{0}})(r_{L}^{2}-r_{R}^{2}) \right]. \nonumber \\
\end{eqnarray}
In other words, one can say FD from phase step is a generalized interference, i.e., superposition or interference of two diffracted wavefronts from a common boundary or PS, which the interference and diffraction effects are coded into Fresnel integrals and $ \varphi $-dependent terms. In Eq.~(\ref{Ip}), we have assumed that the complex reflection coefficients of the left and the right sides of the step are respectively $ \tilde{r}_L= r_L e^{i\varphi_L} $ and $ \tilde{r}_R= r_R e^{i\varphi_R} $ \cite{pedroti}. It should be noted that the amplitude and the phase of the complex reflection coefficients depend on the incident angle and polarization of the light \cite{pedroti}. Besides,
\begin{eqnarray} \label{phase1}
&& \varphi =\frac{4\pi }{\lambda }h\cos \theta +({{\varphi }_{R}}-{{\varphi }_{L}}) ,
\end{eqnarray}
where $ \theta $ and $ \lambda $ stand for incident angle and wavelength, respectively. Recalling that $ {{C}_{0}}+i{{S}_{0}}=\int_{0}^{{{v}_{0}}}{\exp {{(i\frac{\pi {{v}^{2}}}{2})}_{{}}}}dv $ is the Fresnel integral for ${{v}_{0}}={{x}_{0}}\sqrt{2/\lambda R} $ in which  $ x_0 $ is the distance from point $ P_0 $ to the step edge, see Fig. \ref{fig1}(a), and $ R $ is the distance between points $ P_0 $ and $ P $. For the same materials in both side of the step, i.e., $ \tilde{r}_L= \tilde{r}_R=\tilde{r} $, Eq.~(\ref{Ip}) reduces to the following:
\begin{eqnarray} \label{Ip2}
&& \!\!\!\!\!\!\!\!\!\!\!\! {{I}_{(P)}}={{I}_{0}}{{r}^{2}}\left[ {{\cos }^{2}}(\frac{\varphi }{2})+2(C_{0}^{2}+S_{0}^{2}){{\sin }^{2}}(\frac{\varphi }{2})-({{C}_{0}}-{{S}_{0}})\sin \varphi  \right], \nonumber \\
\end{eqnarray}
where, $ r $ is the module of $ \tilde{r} $ and also in which 
\begin{eqnarray} \label{phase2_ref}
&& \varphi =\frac{4\pi }{\lambda }h\cos \theta  ,
\end{eqnarray}
for the reflection mode and
\begin{eqnarray} \label{phase3_trans}
&& \varphi =\frac{2\pi }{\lambda }h{N}'\left[ \sqrt{{{n}^{2}}-{{\sin }^{2}}\theta }-\cos \theta  \right] ,
\end{eqnarray}
for the transmission mode. In Eq.~(\ref{phase3_trans}), $ n=N/N' $ is the ratio of the plate refractive index to the refractive index of the surrounding medium.

It should be noted although here we have used the scaler theory of Fresnel-Kirchhoff integrals to describe Fresnel diffraction from a phase step, but the effects of polarization (s- and p-polarization) of the incident light beam or optical constants is still mapped into the Fresnel coefficient of the materials on the left/right side of the step which are given by \cite{pedroti}
\begin{subequations} \label{Fresnelcoefficient}
\begin{eqnarray}
&&  \!\!\!\!\!\!\!\!\!\!\!\!\!\!\!\!  r_{L(R)}^{(s)}=\sqrt{ \frac{(n_0 cos\theta-u_{L(R)})^2 + v_{L(R)}^2)}{(n_0 cos\theta + u_{L(R)})^2 + v_{L(R)}^2}}, \label{r_s} \\
&& \!\!\!\!\!\!\!\!\!\!\!\!\!\!\!\!  \varphi_{L(R)}^{(s)}=\tan^{-1} \left[ \frac{2 v_{L(R)} n_0 \cos\theta }{u_{L(R)}^2 + v_{L(R)}^2 - n_0^2 \cos \theta} \right] , \label{phi_s} \\
&&  \!\!\!\!\!\!\!\!\!\!\!\!\!\!\!\!  r_{L(R)}^{(p)} \!= \! \sqrt{\frac{ \! \left[ (n_{L(R)}^2 \!-\! k_{L(R)}^2 ) \! \cos\theta \! -\! n_0 u_{L(R)} \! \right]^2 \!\! + \! \left[2 n_{L(R)} k_{L(R)} \cos \theta - n_0 v_{L(R)} \right]^2 }{ \! \left[ (n_{L(R)}^2 \!-\! k_{L(R)}^2 ) \! \cos\theta \! + \! n_0 u_{L(R)} \! \right]^2 \!\! + \! \left[2 n_{L(R)} k_{L(R)} \cos \theta + n_0 v_{L(R)} \right]^2 }}  \label{r_p} , \nonumber \\
&& \\
&& \!\!\!\!\!\!\!\!\!\!\!\!\!\!\!\!  \varphi_{L(R)}^{(s)}=\tan^{-1} \left[\! 2n_0 \cos\theta \frac{2 n_{L(R)} k_{L(R)} u_{L(R)} - (n_{L(R)}^2-k_{L(R)}^2)v_{L(R)} }{(n_{L(R)}^2 + k_{L(R)}^2)^2 \cos^2 \theta - n_0^2 (u_{L(R)}^2 + v_{L(R)}^2)} \right], \label{phi_p} \nonumber \\
\end{eqnarray}
\end{subequations}
where the real functions $ u_{L(R)} $ and $ v_{L(R)} $ satisfy \cite{pedroti}
\begin{eqnarray}
&& (u_{L(R)} + i v_{L(R)})^2= \tilde n_{L(R)}^2 + n_0^2 \sin^2 \theta  ,
\end{eqnarray}
with $ \tilde n_{L(R)}= n_{L(R)} + i k_{L(R)} $ stands for the complex refractive index. Also, it is clear that the effects of the optical constants ($ n_{L(R)}, k_{L(R)}$) of the materials on both side of the step affect the intensity distribution or the phase difference via the complex reflection coefficients.

\begin{figure} [h!]
	\centering
	\includegraphics[width=8cm]{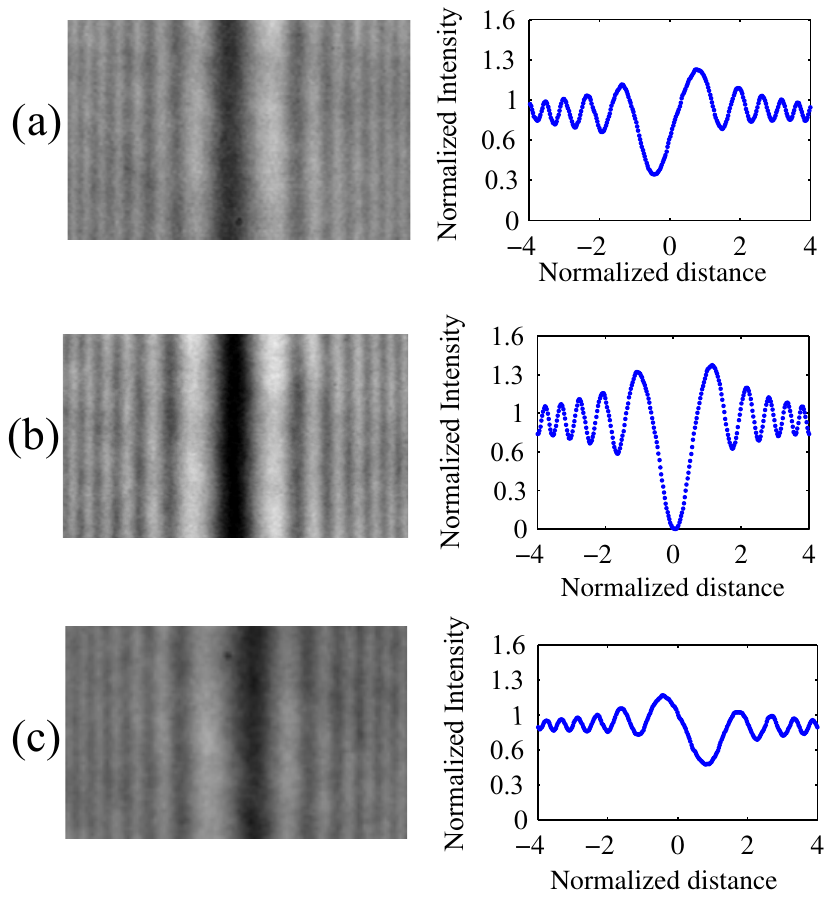}
	\caption{(Color online) Experimental diffraction patterns and their intensity profiles of light beams diffracted from a step of height $ h=450 $nm at incident angle (a) $ \theta \simeq 35^\circ $, (b) $  \theta \simeq 69^\circ $ and (c) $  \theta \simeq 56^\circ $. Vertical axis is normalized intensity that is fraction of intensity in CCD camera to average intensity in far distance from the central dark fringe. The horizontal axis is normalized distance from edge of step that is $v=x \sqrt{2/\lambda R} $, where $ x $ is the distance from the step edge in pixel size of CCD in micrometer.}
	\label{fig2}
\end{figure}

Typical diffraction patterns and their intensity profiles are shown in Fig.~(\ref{fig2}). As the plots show, the fringes' visibility is changed by the change of the step height or incident angle. Also, the plots indicate that the fringes shift slightly to the left or to the right by the change of optical path difference. However, shifts for central fringes are more pronounced, for example in this figure the central minimum shifts from the normalized distance $ V_0 \simeq -0.5  $ to $ V_0=0$ and then to $ V_0 \simeq 0.5 $, respectively from (a) to (c). For a step with similar reflection coefficients on both sides of the step, the visibility defined by \cite{18,19,20,21,22}
\begin{eqnarray} \label{vis}
&& V=\frac{\frac{{{I}_{\max ,L}}+{{I}_{\max ,R}}}{2}-{{I}_{\min ,C}}}{\frac{{{I}_{\max ,L}}+{{I}_{\max ,R}}}{2}+{{I}_{\min ,C}}} ,
\end{eqnarray}
where $ I_{max,R} $, $  I_{max,L} $ and $ I_{min,C} $ stand for the maximum and minimum intensities of the three central fringes, varies between zero and one ($ 0 \le V \le 1 $).

\begin{figure}
	\centering
	\includegraphics[width=6cm]{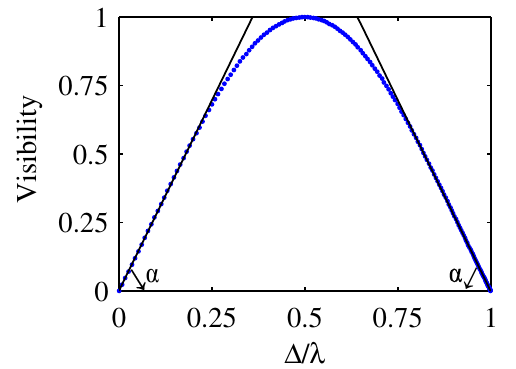}
	\caption{(Color online) Visibility versus optical path difference divided by wavelength, for the three central diffraction fringes of light diffracted from a step with equal reflectance on both sides of the step edge ($ \tilde{r}_L=\tilde r_R=\tilde r $). As shown approximately for visibilities less than $ 0.7 $ the behavior is linear. Theoretically has been shown $\alpha ={{\tan }^{-1}}2.77 $ \cite{22}, therefor the height can be found using relation $ h=s\lambda/5.54 $ \cite{22}, where $ s $ is experimental slope.}
	\label{fig3}
\end{figure}

In Fig.~(\ref{fig3}), the defined visibility, Eq.~(\ref{vis}), is plotted versus optical path difference divided by wavelength, $ \Delta/\lambda $. As can be seen from Fig.~(\ref{fig3}), approximately for visibility between $ 0 $  and $ 0.7 $ ($ 0 \le V \le 0.7 $) two symmetrical lines are fitted on the visibility curve. Thus, by changing the incident angle one can always shift the visibility into interval $ 0-0.7 $ and use the plotted straight lines to deduce the step height. The plot in Fig.~(\ref{fig3}) is a universal curve and one can always get the step height by plotting visibility versus incident angle and equating its slope with the slope of the relevant plotted line in Fig.~(\ref{fig3}), \cite{22}.

Using this technique, i.e., the \textit{visibility} method, the step heights in the range of few tens nanometers up to several millimeters have been measured \cite{22,23}. Also, the visibility change versus step height or incident angle in 1D FD from PS has been applied to the measurements of plate thickness, wavelength, and dispersion relation \cite{22}, nano displacement \cite{24}, refractive indices of solids and liquids \cite{25,26,27}, specification of the temperature profile around a very thin hot wire \cite{28}, refractive index of fiber \cite{29}, spectral modification by singular line \cite{30}, coherence length, correlations and shape of the spectrum \cite{31}, nonlinear refractive index \cite{32} and refractive index of the transparent films \cite{33}, phase step diffractometer \cite{34} and its application to wavemetery \cite{35}, focal/back-focal length measurement \cite{36,dashtdarfocal2}, measurement of thickness of thin film by white light diffractometry \cite{37} and phase singularity \cite{38}. Also, the FD from a step with two different materials on its both sides in the reflection mode has been theoretically formulated \cite{39}. Furthermore, recently the FD from the PS with arbitrarily oriented has been formulated in the general case in reflection and transmission mode \cite{40,41} and it has been shown that the FD in transmission mode have potential to be a hologram which can provide a
3D image of the plate edge \cite{41,dashtdar3D-1,dashtdar3D-2}. Very recently, FD from PS has been used to determination of the spectral line profile of light sources \cite{hasaniline}.

It should be mentioned that the theory which has been considered here and in the other published paper on Fresnel diffraction from phase step is only valid for the abrupt changes in the phase step. As far as we know, it has not been still formulated for the arbitrary surface profile or arbitrary shape of the step. 
However, very recently \cite{fracturedplane} it has been formulated and experimentally verified for the abrupt/sharp change in the derivative of the phase in a Fresnel’s biprism which produces two crossing plane wavefronts (sharp change in orientation of the propagation of wavefronts in a boundary). Therefore, for the other surface profile or for the other shape of the step the present theory cannot be applied and should be reformulated.

Based on the above mentioned investigations, only the visibility method are used for the different measurements up to now. But, the visibility method is not a fast method since it requires the diffraction pattern at \textit{many} angles for high precision measurements \cite{21,39}. That is why we would like to introduce the \textit{``fitting"} method as a \textit{fast} method in the next subsection which can only use the diffraction pattern at \textit{one} angle and thus leads to thickness measurement with high accuracy. 


\subsection{ Fitting on the simulated intensity profile of a step with two different materials} \label{ellipsometry}

Now, we are going to explain the fit mechanism \textbf{through} simulation which illustrates how by ``\textit{fitting}" the theoretical intensity distributions on the corresponding experimental intensity profiles, one can measure the step height $ h $ with high precision when both sides of the step has the same material as is explained in the next section [equal reflectance on both sides of the step edge, i.e., ($ \tilde{r}_L = \tilde r_R $)]. 

Let us illustrate the \textit{ability} of the fitting method. For clarity, we consider a general step with two different materials on either side [unequal reflectance on the sides of the step edge, i.e., ($ \tilde{r}_L \neq \tilde r_R $), for example two metals] and show that by fitting the theoretical intensity profile i.e., Eq.~(\ref{Ip}), on the \textit{simulated} intensity profile, the required data for determining the step height and the optical constants of the materials are achievable. Nevertheless, in the experimental part we only focus on a step with the same material on either side for thickness measurement,i.e., thickness measurement.

\begin{figure}
	\centering
	\includegraphics[width=9cm]{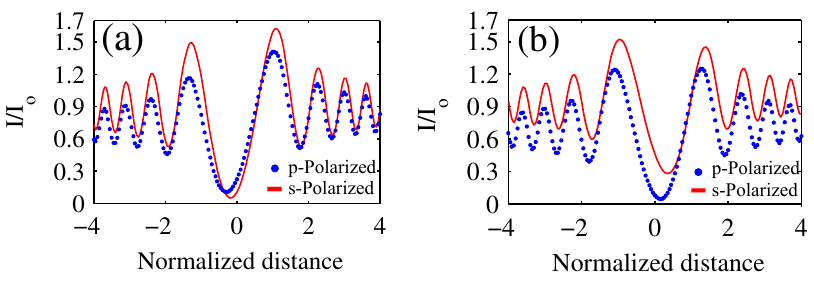}
	\caption{(Color online) The theoretical simulated intensity profiles on the diffraction patterns of light diffracted from a step of height $ h=300 $nm coated with Al on the left and with Cu on the right sides of the step edge, illuminated by s and p polarized lights of wavelength $ \lambda=589.3 $nm (Yellow, Na Lamp) at incident angle (a) $ 55^\circ $ and (b) $ 76^\circ $. Solid and dotted curves are referred to s polarization and p polarization, respectively.}
	\label{fig4}
\end{figure}

In our \textit{simulation}, we consider as a general example a step with height 300 nm coated with Cu and Al on the right and the left sides of the step edge, respectively. In order to \textit{simulate} the intensity distributions on the diffraction patterns of light diffracted from the latter step we have substituted the optical constants of Cu and Al for $ \lambda=589.3 $nm, namely, $ n_{Al} \approx 1.16 $ and $ k_{Al} \approx 7.2 $, and  $ n_{Cu} \approx 0.47 $ and $ k_{Cu} \approx 2.81 $ \cite{42}, in Eq.~(\ref{Ip}) and have plotted the function for two values of incident angles $ 55^\circ $ and $ 76^\circ $ for s- and p-polarizations (see Fig.~(\ref{fig4})). The solid and the dotted curves correspond to p and s polarizations, respectively. Note that the polarization is taken into account by considering Eqs.~(\ref{r_s})-(\ref{phi_p}) in Eq.~(\ref{Ip}). 
Every point of these plots provides an equation that involves the step height and the optical constants of the step materials. Thus, fitting the theoretical intensity distribution, Eq.~(\ref{Ip}), on the corresponding simulated intensity profile of s- and p-polarization at each angle is equivalent to solving a large number of independent equations with few unknown parameters, i.e., $ h $ and $ r_L, \varphi_L, r_R, \varphi_R $ which themselves depend on the incident angle $ \theta $ and optical constants $ n_{L,R} $ and $ k_{L,R} $ \cite{pedroti} via Eqs.~(\ref{r_s})-(\ref{phi_p}). Also, it should be remembered that, away from the step edge at each side, the reflection coefficient is determined by the optical constants of the material of that side. Therefore, by analysis of the intensity profile of the diffracted light, one can directly evaluate the ratio of the reflection coefficients of the two sides. Thus, the optical constants obtained by the fitting technique can be controlled by the latter ratio. In order to test the approach in our simulation, we tried to deduce the optical constants and the step height used in plotting the simulated curves in Fig.~(\ref{fig4}). We have evaluated Eq.~(\ref{Ip}) with unknown values of optical constants and step height. However, the best fitting occurs for the values that were used in the plots of Fig.~(\ref{fig4}) with satisfactory precision about $ 0.1\% $. The reason for this high accuracy in the simulation in spite of the experimental case is that we have ignored the effects of noise in our simulations.

Surprisingly, as will be shown in our experimental part for the thickness measurement, a step with the same material on its sides ($ \tilde r_L =\tilde r_R=\tilde r $), fitting the theoretical intensity distributions (in this case Eq.~(\ref{Ip2})) on the corresponding experimental intensity profiles occurs with an uncertainty less than few nanometers while using modest equipment (see table \ref{tab}). It should be noted that the fitting method is very fast in comparison to the visibility method because we can use a diffraction pattern at \textit{one} angle and obtain high precise values for unknown thicknesses or step height. Thus, the technique has a good potential to become a highly valuable \textit{fast}-method for thickness measurement of reflective thin films.

\section{Experimental procedure and results} \label{experiment}

Here, we apply the fitting technique to measure \textit{film thickness} when both sides of the step have the same material (same reflectance on both side of the step ($ \tilde r_L =\tilde r_R=\tilde r $) and compare the results with the thicknesses measured through the visibility curve corresponding to the three central fringes of diffraction pattern and also with the other techniques.

At first, we prepared steps of different heights by Dc-sputtering coating Al and Cr on glass slides that were partly masked. Then, we removed the masks and coated the entire slides with Al or Cr in order to have steps with the same reflectance on the both sides. The reported heights which have been obtained by the piezo crystal during the coating process are $ h \sim 50 $nm, $ h \sim 150 $nm, and $ h \sim 450 $nm.

In the following, we explain with more details how we have prepared the physical step by coating and masking (photolithography) in order to generate an abrupt change in the step. 
First, we coat a target layer (for example a reflective layer like Al and Cr or even a dielectric layer such as MgF2 which we want to measure its thickness), on a slide as a substrate which is partly masked for example by a soda lime or anything having a sharp edge. Thus, when we remove the mask we have a sharp step with a height equal to the thickness of coated target layer. Then, we again coat the entire slide with a high reflective layer as a second layer such as Al, Cr or something like that in order to have the same reflectance on both side of the step. Moreover, instead of masking we can etch the target layer through photolithography in order to make a sharp step and then coat the entire slide by a high reflective second layer. In this manner, it is clear that the height of the step is equal to the thickness of coated target layer (first layer) and also the step has a physical abrupt change (thus our theory is still valid for this case).
Furthermore, as far as we know, the second layer coating, i.e., reflective layer, does not change neither the step height caused by the target layer nor its shape or its sharpness. In other words, after coating the second layer in regions where the step exists, it becomes similar to the original step caused by the target layer and also has an abrupt change, i.e., sharp step edge. Therefore, we are sure that our theory works well and is valid in this case. However, in this manner our sample preparation method may be seen as a destructive one because in order to have a sharp step edge, which has an abrupt change, the surface of the original target layer may be affected by a high reflective second layer coating or masking or lithography.

\begin{figure}
	\centering
	\includegraphics[width=8.5cm]{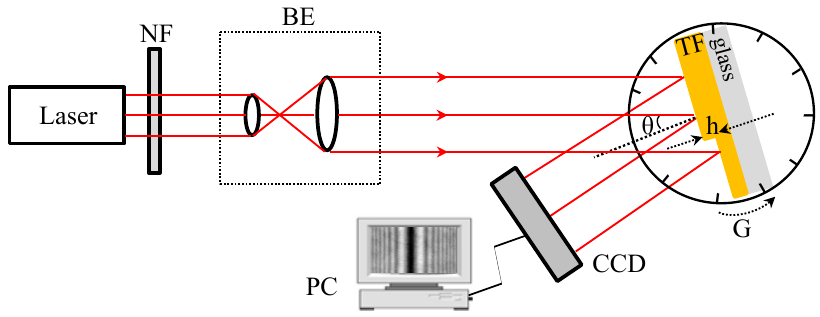}
	\caption{(Color online) The sketch of the experimental setup used for measuring film thickness. The laser beams after passing through neutral filter NF is expanded by beam expander BE. The expanded beam strikes the sample mounted on goniometer G and the diffracted beam is captured by CCD and fed to PC.}
	\label{fig5}
\end{figure}

The scheme of the experimental set up is shown in Fig.~(\ref{fig5}). The He-Ne laser beam with adjustable power with maximum output power $ 5 mW $, after passing through neutral filter NF, is expanded by beam expander BE with maximum beam divergent angle of $ 10^{-7} $rad and beam diameter $ 2cm $. The expanded beam strikes the step mounted on motorized goniometer G that can rotate with precision of about one arc minute (with maximum rotation resolution $ 48 $ arc second). An astronomical ultra linear CCD Camera (Celestron Skyris 618M Mono CCD Camera) with 5.7 µm pixel size which provides 12-bit image output is mounted on an arm that can turn around the axis of the goniometer, receives the diffraction patterns and feeds them to a PC. Note that NF is used for the decreasing the intensity in order to prevent saturation of the CCD. Before starting, we check the linearity response of our CCD by Malus's law. By rotating the sample, the incident angle is varied and diffraction patterns are recorded at desired incident angles. For removing the effects of the background in each angle we record the image of background in CCD when the diffracted beam is blocked and then omit it from the original intensity profile of the diffracted light. 
Moreover, since the reflectance of the both side of the step are the same, because of the same material on both side of the step, we illuminate the step with unpolarized light. Because, in this case without knowing the optical constants of the step one can determine the thickness only by knowing the reflectance which can be easily obtained from the experimental intensity profile [in next paragraph we explain it with more details]

It should be mentioned that since in our experiment the materials on both sides of the step are the same, there is no phase difference due to the optical constants of the materials, i.e., $ \varphi_L^{(s(p))} = \varphi_R^{(s(p))} $ in Eqs.~(\ref{Ip}) and (\ref{phase1}). Thus the total phase difference is only due to the step height or thickness of thin film, i.e., $ \varphi= (4\pi/\lambda) h \cos \theta $. 
Moreover, although the amplitude of reflection coefficients on both side of the step are the same ($ r_L^{(s(p))} =  r_R^{(s(p))} =r_{s(p)}$), the effect of polarization or the optical constants still remains in Eq.~(\ref{Ip}) through the reflection coefficient of the surface, i.e., $ I_0  r_{s(p)}^2 $ [see Eq.~(\ref{Ip2})]. 
Thus, by using the normalized ($I/ I_0  r_{s(p)}^2 $) experimental intensity profile in all experimental figures we don’t need to know the polarization of light or the optical constant of the material on both side of the step in our fitting process for measuring the thickness of thin films.

Furthermore, it should be mentioned that in this case (step with the same material on both side of the step) we illuminate the step by unpolarized light and then use the normalized intensity profile ($ I/I_0 r^2 $). It is worth to mention that in order to normalize the experimental intensity, one can experimentally determine the reflectance of the unpolarized incident light [$ I_0  r^2 $ with reflection coefficient $ r $] away from the step edge or central fringes. 
Therefore, it is obvious that in this case we don’t need to know the optical constants and polarizations. So, the only unknown parameter in our fitting process is the step height $ h $ or film thickness which can be obtained from the fit.

\begin{figure}
	\centering
	\includegraphics[width=8.5cm]{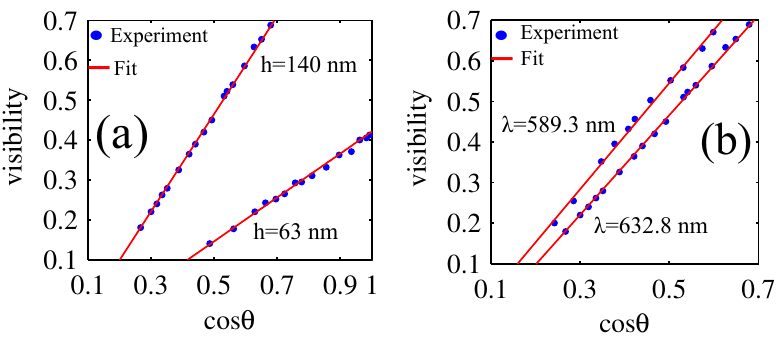}
	\caption{(Color online) The dots are experimental visibility of the three central diffraction fringes versus the cosine of incident angle, $ \cos \theta $, for steps of height (a) $ h=63$nm and $ h=140 $nm illuminated by a light beam of wavelength $ \lambda=632.8 $nm, (b) for a step of height $ h=140 $nm illuminated by light beams of wavelengths $ \lambda=589.3 $nm and $ \lambda= 632.8 $nm. The lines are the best fitted curves. Here, the values of $ h $ is obtained by the visibility technique.}
	\label{fig6vis}
\end{figure}

The dots in Fig.~(\ref{fig6vis}) represent the experimental visibilities, defined by Eq.~(\ref{vis}), versus the cosine of incident angle for steps of heights (a) $ h=63 \pm 6 $nm, (b) $ h=140 \pm 5 $nm illuminated by lights of the wavelength $ \lambda=632.8 $nm [here, heights are obtained by the visibility technique]. The dots in Fig.~\ref{fig6vis}(b) are experimental visibilities for a step of height $ h=140 $ illuminated by lights of wavelengths $ \lambda=632.8 $nm and $ \lambda=589.3 $nm. The slopes of the best fitted lines provide the step heights \cite{21,22}. In Fig.~\ref{fig6vis}(a), as the plot indicated by $ h=63 $nm shows about 15 experimental spots have been used to draw the best fitted line for a film of thickness $ h=63 $nm. This clearly indicates that the method allows measuring much lower thicknesses. In addition, use of plot for the evaluation of thickness improves the reliability and the accuracy of the results. 

For obtaining the steps height by \textit{fitting method }we try to fit the function 
\begin{eqnarray} \label{I_fit}
&& \!\!\!\!\!\!\!\!\!\!\!\! I_f(x,\theta, \lambda)= R_r \left[ {{\cos }^{2}}(\frac{\varphi}{2}) \right. \nonumber \\
&&  +2\left(C^2(a x +V_0) + S^2(a x +V_0)\right){{\sin }^{2}}(\frac{\varphi }{2}) \nonumber \\
&&  \left. - \left(C(a x +V_0)-S(a x +V_0) \right) \sin \varphi \quad \right] , 
\end{eqnarray}
on the experimental intensity profiles. $ C,S $ are the Fresnel integrals. In Eq.~(\ref{I_fit}) $ \varphi=\varphi(\theta,\lambda;h)=\frac{4\pi }{\lambda }h\cos \theta $ which includes $ h $. $ R_r $, $ a $, $ V_0 $ and $ h $ are the parameters should be obtained by fitting. Note that $ R_r= I_0 r^2$ is normalization coefficient or reflectance of the unpolarized incident light which is used to normalize intensity, $ V=a x $ is the normalized distance to step edge, $ V_0 $ stands for normalized distance from central minimum to the step edge and $ h $ is the step height or thickness of thin film. Also, the horizontal position in the experimental diffraction pattern $ x $, incident angle $ \theta $ and wavelength $ \lambda $ are specified in each experimental recorded pattern. As is clear, by knowing the reflectance of the unpolarized incident light, $ R_r $, as we explained we don't need to know the optical constants, and thus we can directly measure the thickness $ h $ from fit without knowing the optical constants

\begin{figure*}
	\centering
	\includegraphics[width=15cm]{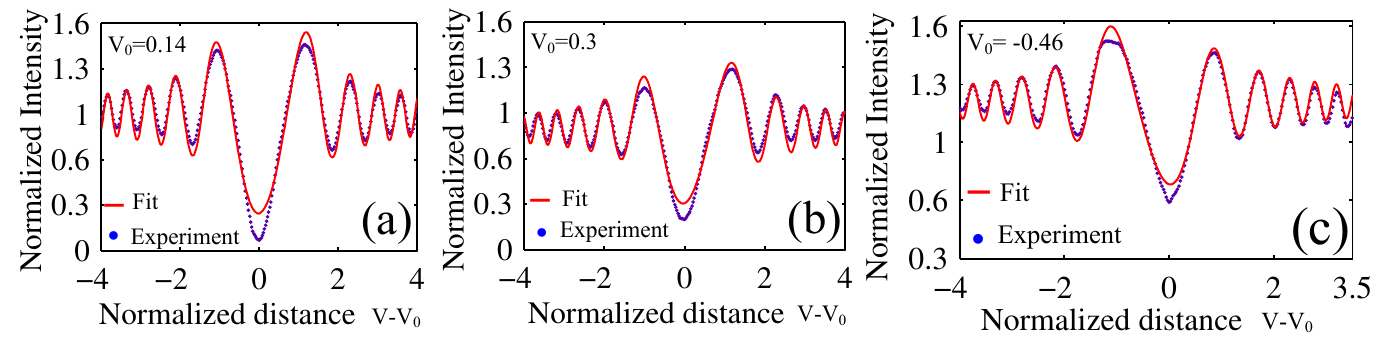}
	\caption{(Color online) The dots represents the experimental intensity profiles on the diffraction patterns of He-Ne laser ($ \lambda=632.8 $nm) beam diffracted from a step of height (a) $ h=140 $nm, (b) $ h=448 $nm, and (c) $ h=63 $nm at incident angles $ 29^\circ $, $ 30^\circ $ and $ 7^\circ $ respectively. The continuous curves are the theoretical curves that have been fitted for heights (a) $ h=141 \pm 5 $nm, (b) $ h=448 \pm 3 $nm, and (c) $ h=63 \pm 4 $nm. $ V $ is normalized distance to step edge and $ V_0 $ is normalized distance from central minimum to the step edge which is obtained from the fitting process. }
	\label{fig7}
\end{figure*}

In Fig.~(\ref{fig7}), the dots represent the experimental intensity profiles on the diffraction patterns of lights diffracted from the steps used in the visibility technique, at incident angles (a) $ \theta= 29 ^\circ $, (b) $ \theta=30^\circ $ and (c) $ \theta=7^\circ $ for $ \lambda=632.8 $nm. The solid lines are the best fitted theoretical curves that have occurred at the step height (a) $ h=141 \pm 5 $nm, (b) $ h=448 \pm 3 $nm and (c) $ h=63\pm 4 $nm [here, the values of $ h $ are obtained by ``fitting" technique]. According to the plots in Fig.~(\ref{fig7}), fitting at central minimum and maxima shows some deviation. This is observed almost in all cases. The slow change of intensity at the maxima or minima of intensity profile and lack of ideal sharpness at the step edge could be the reason.

\begin{figure}
	\centering
	\includegraphics[width=7cm]{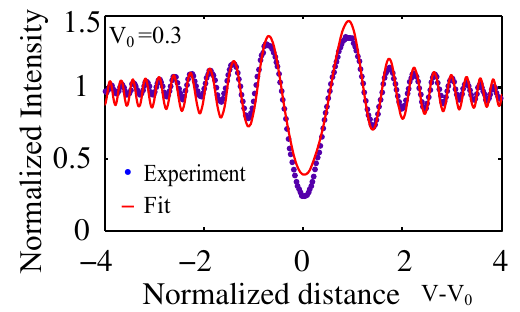}
	\caption{(Color online) The dots represent the experimental intensity profile on the diffraction pattern of a beam of Na source ($ \lambda=589.3 $nm) diffracted from a step of height $ h=140 $nm at incident angle $ \theta= 51^\circ $. The solid is the best fitted curve at $ h=142 \pm 7 $nm. $ V $ is normalized distance to step edge and $ V_0 $ is normalized distance from central minimum to the step edge which obtained in fit. }
	\label{fig8}
\end{figure}

In Fig.~(\ref{fig8}), the dots represent the experimental intensity profile on the diffraction pattern of light diffracted from the step of height $ h=140 $nm, obtained by the visibility technique, for $ \lambda=589.3 $nm (Na Lamp). The best fitting has occurred for height $ h=142 \pm 7 $nm.

Using both methods, i.e., the visibility and fitting, the described experiments were repeated for different samples and wavelengths. Some of the results are presented in Table~(\ref{tab}). As the table shows the results of both techniques are in excellent agreement with each other. As is evidence from Table~(\ref{tab}), the precision of the measurements is clearly higher than what is obtained by the conventional interferometry methods \cite{43}. The main reason for higher precision is that in the Fresnel diffraction from step the quadratic term of the phase change is also accounted.

\begin{table*} 
	\centering
	\caption{\bf Film thicknesses that were measured by applying Fresnel diffraction of light to steps fabricated by coating. Comparing the three last column shows that the results are in good agreement. Also, it shows that the fitting technique could be a fast and high precision method for measuring the film thicknesses. }
	\begin{tabular}{ccccc}
		\hline
		Wavelength & Sample & Thicknesses (nm) & Thicknesses (nm) & Reported Thicknesses (nm)  \\
		$ \lambda $ (nm)&  & by fitting & by visibility  & by piezo crystal in coating  \\
		\hline
		$632.8$ & Cr & $ h=448 \pm 3 $ & $ h=448 \pm 9  $ & $ h \sim 450 $\\
		$632.8$ & Al & $ h=141 \pm 5 $ & $ h=140 \pm 5  $ & $ h \sim 150 $ \\
		$632.8$ & Cr & $ h=63  \pm 4 $ & $ h=63  \pm 6  $ & $ h \sim 50 $ \\
		$589.3$ & Al & $ h=142 \pm 7 $ & $ h=139 \pm 9  $ & $ h \sim 150 $ \\
		\hline
	\end{tabular}
	\label{tab}
\end{table*}

It should be mentioned that for an easy thickness measurement in our experiment we have written and developed a software in MATLAB and Python. It can control our motorized goniometer G in order to rotate the sample/CCD and record the diffraction pattern and then send it to the PC for analysis. The required time for the measurement using the visibility and the fitting methods are respectively about less than 5 minutes and about one second. This clearly shows that the fitting method is much \textit{faster} than the visibility method because in fit method we are able to measure the thickness with high precision only by using a \textit{one} recorded diffraction patterns at \textit{one} \textit{angle} while for a good measurement by visibility we should use the diffraction patterns at\textit{ many angles} and it takes too much time.

In Table~(\ref{tab2}), we compare the thicknesses measured for the other samples by FD from phase step to some other known methods such as Profilometer (Dektak or Tullystep or Stylus profilometers), AFM and Interferometer. 
As shown in Table~(\ref{tab2}), the results are in good agreement. This indicates that the FD method is trustworthy and comparable to other methods in spite of its simplicity.

\begin{table*}
	\centering
	\caption{\bf Comparison between thickness measurement by FD at He-Ne wavelength and other methods such as Profilometer (Stylus profilometer, Dektak or Tullystep), AFM and Interferometer. }
	\begin{tabular}{cccccc}
		\hline
		Sample & Thickness (nm)   & Thicknesses (nm) & Thicknesses (nm) & Thicknesses (nm)  & Thickness (nm) \\
		number & by coating equipment& by Profilometer  & by AFM           & by Interferometer & by Fresnel Diffractometer \\
		\hline
		1      & 100               & 80               & 60                & 68                & 66  \\
		2      & 200               & 200              & ---               & 206               & 208  \\
		3      & 300               & 340              & 320               & 335               & 328  \\
		4      & 400               & 366              & ---               & 372               & 380 \\
		\hline
	\end{tabular}
	\label{tab2}
\end{table*}

\begin{figure}
	\centering
	\includegraphics[width=8.5cm]{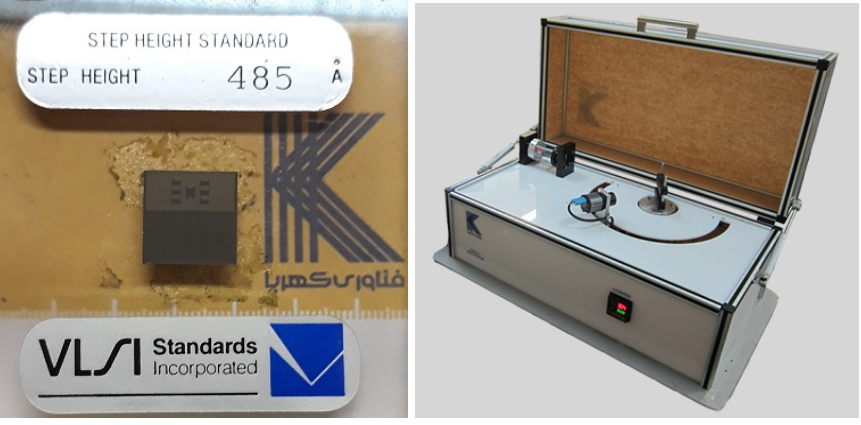}
	\caption{(Color online). (Left panel) VLSI Standard step height with the steps hight $ 485 $A$\! ^\circ $ \cite{VLSI} which we have used for testing the accuracy of the fitting and visibility method based on Fresnel diffraction from nano phase step. (Right panel) The picture of our manufactured device,FDP-D1, based on FD from PS \cite{kahroba} for measurement of thickness of thin films.}
	\label{figVLSI}
\end{figure}

Finally, let us answer to an important question that how much FD from phase step is really accurate. For answering this question, we have tried to measure the VLSI Step Height Standard with the step height $ 485 $A$\!^\circ $ (VLSI Standards Incorporated \cite{VLSI}). Fig.~(\ref{figVLSI}) shows our VLSI standard step which have been used here. This Step Height Standard consists of some etched patterns and some steps height in a quartz substrate and are usually used to calibrate optical or stylus based profilometers \cite{VLSI}. 
We have used one of its standard step in order to test the accuracy of our method, i.e., fitting and visibility methods based on FD from phase step. We have applied both visibility and fitting techniques in order to measure the step height. 
Finally, we have obtained the standard step height as $ (50 \pm 3) $nm and $(52 \pm 4) $nm which are respectively due to the measurement by fitting and visibility techniques; and coincidence to VLSI Step Height Standard. This shows that the thickness measurement by FD from PS (both methods of visibility and fit) , is not only comparable and competitive but also trustworthy and accurate in real measurements.

\section{Conclusion, discussion and outlook } \label{con}
This study shows that FD from phase steps is a rich subject with many metrological applications. The application of the \textit{fitting} technique to film thickness measurement is easy, reliable, fast and precise and can be applied in a very wide range of thicknesses with modest instrumentation.

It should be noted that we have manufactured and trademarked a device (Fresnel diffractometer), FDP-D1 which is shown in right panel of Fig.~(\ref{figVLSI}), for measurement of thickness of thin films based on FD from the nano PS (using both fitting and also visibility methods) \cite{kahroba}.

\begin{figure*} 
	\centering
	\includegraphics[width=15cm]{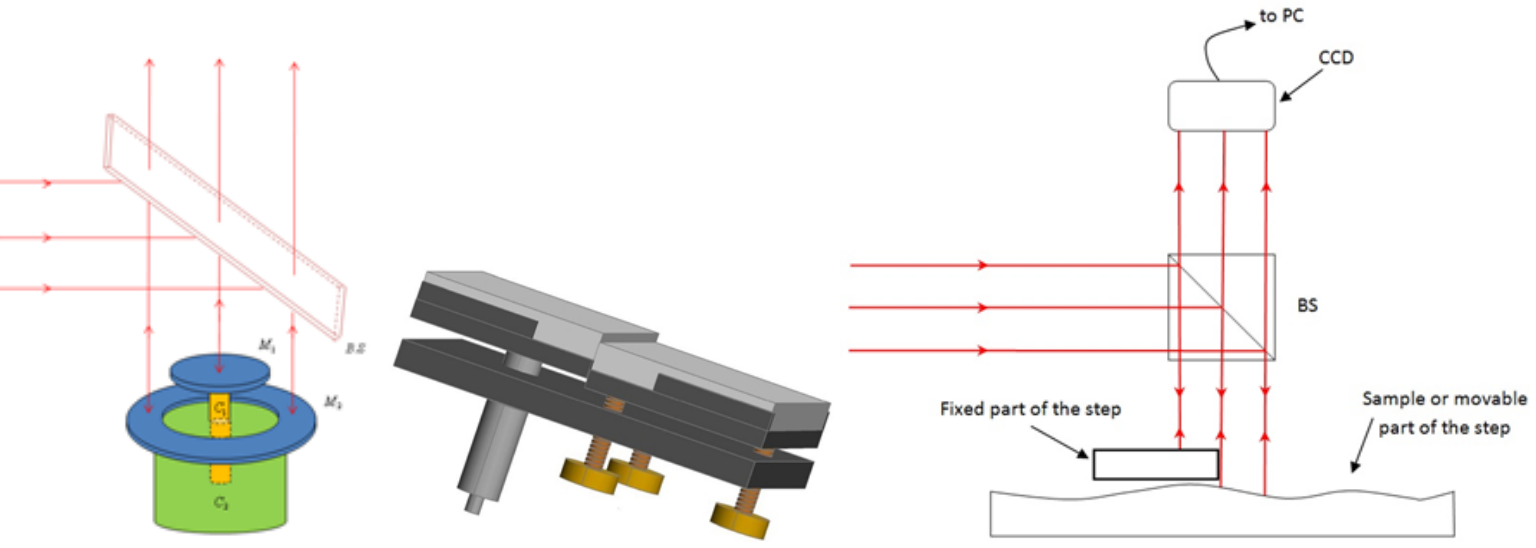}
	\caption{(Color online) (Left panel) A circular phase step with variable height which has been introduced in Ref.~\cite{22} and can be used in order to manufacture a nanopositioner based on FD from 2D PS. (Central panel) A typical 1D phase step with variable height \cite{34,35} which can be used to  manufacture a wavemeter, or spectrometer. (Right panel) A sketch of the device for surface topography using FD from 1D PS in reflection mode.}
	\label{fig10}
\end{figure*}

We have mentioned many metrological applications of FD from PS during these two decades in subsection~2\ref{ellipsometry} \cite{18,19,20,21,22,23,24,25,26,27,28,29,30,31,32,33,34,35,36,dashtdarfocal2,37,38,39,40,41,dashtdar3D-1,dashtdar3D-2,hasaniline,fracturedplane}. Nevertheless, it is worth here to point out the importance of FD and its potential in other applications.  Let us firstly explain the importance of FD from PS. Metrological applications of interferometry are numerous and well established. In interferometry, changes in a physical quantity are converted into changes of optical path difference (PD) between the interfering waves, and then the behavior is studied. This approach has been applied to the measurements and studies of many physical properties. While the conventional FD doesn't permit such conversion, FD from a phase step permits changing PD between the waves diffracted from the two sides of the step edge. This possibility provides reliable measurements of many physical quantities with the following significant advantages compared with interferometry:

(a) In FD from a PS the intensity at each point of the diffraction pattern provides information about the phase change which considerably improves the precision of the measurements.

(b) The PD is variable by changing the step height and light incident angle. The latter provides very smooth change of PD.

(c) The technique can be applied in reflection and transmission modes.
	
(d) Since, there is no beam splitter or optical element between the phase step and the fringe detector (CCD); the technique can be applied in a very wide range of wavelengths.
	
(e) The  criteria for our measurements are change of visibility in diffraction fringes with specific shapes and sizes and also the intensity profile distribution which the latter has many information at each angle; therefore, the repetition of the fringes can be clearly traced and  also the intensity noise is low. Moreover, it should be mentioned that thickness measurement using conventional interferometry is based on fringes displacement and therefore is limited to measurement of thicknesses smaller than half-wavelength. For solving this problem one should use two-wavelength interferometry \cite{twowavelength} or white light interferometry \cite{white1}. While FD from PS overcomes this challenge and can be applied for measuring the thicknesses in very large range from nanometer to millimeter or even higher with less complexity (see Refs.~\cite{22,23,24,25,26,33,34,35,37} in addition of the present paper).
	
(f) The manufactured devices based on FD from PS are very compact with very low mechanical noise compared with an ordinary interferometer; therefore, the technique can be utilized in rather rough conditions. Also, as have been explained the devices based on FD from phase step are much more inexpensive than other metrological devices since they use modest, simple and conventional optical setups and elements.
	
(g) According to Eqs.~(\ref{Ip}) and (\ref{Ip2}), there is a sinusoidal term in the expression of the intensity distribution which allows distinguishing the sense of PD.

Here, we point out some potentials of FD from PS in other applications which can be mentioned as outlooks. 
The sketches of two useful designs of 1D FD from PSs are shown in central and right panel of Fig.~(\ref{fig10}). Moreover, a sketch of 2D FD from PS is shown in left panel of Fig.~(\ref{fig10}).
(i) Using a circular step with variable height [see left panel in Fig.~(\ref{fig10})] which have been led to displacement measurement with nanometer accuracy \cite{22}, one can manufacture a nanopositioner or can be used as a feedback for characterizing the piezoelectric nanopositioner. 
(ii) The design in central panel in Fig.~(\ref{fig10}), 1D PS diffractometer, includes two flat mirrors mounted on two independent supports. The mirrors are aligned by the provided screws and one of the mirrors can be displaced perpendicular to its surface in order to change the step height. The step is illuminated in a way that the step edge remains in the incident plane. This design has been patented \cite{34}, fabricated and applied to accurate measurements of wavelength (wavemetry), spectral line width, and specification of spectral line shape \cite{31,34,35}. It can be used to manufacture a spectrometer or wavemeter based on FD from 1D PS which works easy and also is simple and inexpensive.
(iii) The design in right panel of Fig.~(\ref{fig10}) can be used for surface topography and measurement of surface curvatures in a wide range of sizes. Also, any physical effects which affect the curvatures of surfaces can be studied or investigated.
In this device, the step consists of a flat mirror held fixed over the sample surface at a small distance. The set is illuminated perpendicularly in the neighborhood of the plate edge and the reflected diffraction pattern is fed to a CCD connected to a PC. By displacing the sample in two perpendicular directions the surface is scanned and height distribution is evaluated. From theoretical point of view, the wave that is reflected from the surface of the fixed part of the step, the mirror, is the reference wave and the wave reflected from sample surface is the unknown wave.
(iv) Based on the obtained results in Refs.~\cite{40,41,dashtdar3D-1,dashtdar3D-2}, it seems that FD from PS can be utilized in phase microscopy in order to study of biological tissues or inhomogeneity in transparent materials. In other words, one can manufacture a high resolution phase microscope based on FD from phase objects/steps.
(v) It is worthwhile to mention that the measurements and studies based on FD from phase steps/objects have good potentials to be applied in a very large interval of wavelengths, X to IR, and also to be extended to electron and neutron beams to study sub-nanometer steps.

Finally, as an important outlook of the presented \textit{fitting} method in this paper it can provide an efficient and powerful technique for measuring the optical constants of materials in future. However, it should be noted that when both sides of the step are not the same, for example the simulated case in subsection 2\ref{ellipsometry}, the visibility curve has not the universal shape \cite{20,21,39}, and thus we cannot use the visibility method for the measurement of optical constants. However, based on our numerical simulation which is shown in subsection~2\ref{ellipsometry} and also based on our experimental results which are shown in Tables~\ref{tab} and \ref{tab2} [which show that fitting method can provide thickness of thin films precisely], one can conclude that the fitting method based on FD from nano PS has the potential to become a highly valuable method for measuring the optical constants of materials, i.e., Ellipsometry based on FD from PS.

\section*{acknowledgment}

The authors would like to express their gratitude to the referees whose valuable comments have improved the paper substantially. Besides, they acknowledge the staffs in Micro-Electronics Center of Sharif University Branch, Eng. S.~M.~Barzin and Eng. V.~Masahati, for providing some of the samples and they sincerely appreciate their contribution. Also, the authors thank the University of Tehran Science and Technology Park, and Iran nanotechnology initiative council (INIC) for their supports. Furthermore, A.~M.~F. thanks Prof. M.~T.~Tavassoly for his useful comments and discussions to improve the paper and also Dr. A.~Dalafi for helping him in rewriting the paper.

\bigskip


\end{document}